\documentclass[%
 reprint,
 superscriptaddress,
preprintnumbers,
nofootinbib,
 amsmath,amssymb,
 aps,
]{revtex4-1}
\usepackage{amsmath}
\usepackage{amssymb}
\usepackage{amsthm}
\usepackage{MnSymbol}
\usepackage{nicefrac}
\usepackage{amsfonts}
\usepackage{dsfont}

\usepackage[markup=nocolor]{changes}

\usepackage{graphicx}

\usepackage[normalem]{ulem}

\usepackage{dcolumn}
\usepackage{bm}

\usepackage{xcolor}
\usepackage{slashed}
\usepackage{empheq}
\usepackage{hyperref}
\newcommand{\bea}{\begin{eqnarray}}
\newcommand{\eea}{\end{eqnarray}}
\newcommand{\be}{\begin{equation}}
\newcommand{\ee}{\end{equation}}

\definecolor{darkgreen}{rgb}{0,0.4,0}

\newcolumntype{N}{@{}m{0pt}@{}}
\newcolumntype{M}[1]{>{\centering\arraybackslash}m{#1}}
\begin{document}
\author{Gustavo P. de Brito} 
\email{gustavo@cp3.sdu.dk}
\affiliation{CP3-Origins, University of Southern Denmark, Campusvej 55, DK-5230 Odense M, Denmark}
\affiliation{Centro Brasileiro de Pesquisas F\'isicas, Rua Dr. Xavier Sigaud 150, 22290-180 Rio de Janeiro, RJ, Brazil}
	
\author{Astrid Eichhorn} 
\email{eichhorn@sdu.dk}
\affiliation{CP3-Origins, University of Southern Denmark, Campusvej 55, DK-5230 Odense M, Denmark}

\author{Marc Schiffer} 
\email{schiffer@thphys.uni-heidelberg.de}
\affiliation{Institut f\"{u}r Theoretische Physik, Universit\"{a}t Heidelberg, Philosophenweg 16,
			 69120 Heidelberg, Germany}
 
\title{Light charged fermions in quantum gravity}

\begin{abstract}
We study the impact of quantum gravity on  a system of chiral fermions that are charged under an Abelian gauge group. Under the impact of quantum gravity, a finite value of the gauge coupling could be generated and in turn drive four-fermion interactions to criticality. We find indications that the gravity-gauge-fermion interplay protects the lightness of fermions for a large enough number of fermions. On the other hand, for a smaller number of fermions, chiral symmetry may be broken, which would be in tension with the observation of light fermions. 
\end{abstract}

\maketitle
\section{Introduction: Light fermions as a test of quantum gravity}
The observation of light fermions in our universe motivates studies of the interplay of chiral fermions with quantum gravity. Chiral symmetry protects fermions from acquiring masses; conversely, its breaking by quantum gravitational effects would be expected to result in masses of the order of the Planck mass. The presence of light fermions can therefore be used to derive constraints on quantum gravity, as suggested in \cite{Eichhorn:2011pc}.

The question of whether a global chiral symmetry remains unbroken in a given approach to quantum gravity has been answered to varying degrees in different approaches. The results in \cite{Eichhorn:2011pc}, when interpreted in the light of an effective-field-theory setting, where a finite new-physics scale is present, suggest that quantum gravitational fluctuations do not necessarily trigger chiral symmetry breaking.
In comparison, general arguments for the breaking of global symmetries through quantum gravity have been substantiated in string theory  \cite{Banks:1988yz,Kamionkowski:1992mf,Kallosh:1995hi,Banks:2010zn} and in the context of the AdS/CFT conjecture \cite{Harlow:2018jwu}, specifically related to the weak-gravity conjecture  \cite{ArkaniHamed:2006dz}, see also \cite{Palti:2019pca} for a review; although to the best of our knowledge not specifically for the case of a chiral global symmetry.
On the lattice, the  fermion doubling problem, formalized in the Nielsen-Ninomiya theorem \cite{Nielsen:1980rz,Nielsen:1981xu}, implies that at finite lattice spacing, chiral fermions cannot exist and the continuum limit has to be taken in a careful fashion to recover them. While discrete quantum-gravity approaches are not regular lattices, the theorem could nevertheless be viewed as a hint at the potential difficulty to accommodate chiral fermions. As a concrete example, within Loop Quantum Gravity, this question has been discussed in the literature~\cite{MontesinosVelasquez:1997ht,Barnett:2015ara,Gambini:2015nra}. Other discrete models, such as causal sets~\cite{Surya:2019ndm}, depart from a manifold-like structure to that extent that currently the very definition of a fermion field, chiral or not, is an open question. In contrast, in Euclidean Dynamical Triangulations, progress has recently been made using K\"ahler fermions~\cite{Catterall:2018dns}. In a first lattice study, a global U(1) symmetry that is related to chiral symmetry in the continuum limit, appears unbroken~\cite{Catterall:2018dns}.

Within the functional Renormalization Group (FRG) \cite{Wetterich:1992yh, Ellwanger:1993mw,Morris:1993qb,Reuter:1993kw} approach to asymptotic safety, pioneered in \cite{Reuter:1996cp}, the interplay of quantum gravitational fluctuations with chiral symmetry was first explored in \cite{Eichhorn:2011pc}, see also \cite{Meibohm:2016mkp,Eichhorn:2017eht}. Despite the attractive nature of the classical gravitational interaction, quantum gravity fluctuations do not appear to lead to bound-state formation and chiral symmetry breaking. This is in contrast to non-Abelian gauge theories, where chiral symmetry breaking in was studied with the FRG in~\cite{Gies:2001nw,Braun:2005uj,Braun:2006jd,Braun:2006zz,Braun:2008pi,Haas:2010bw,Braun:2010vd,Braun:2011pp}. At the technical level, this difference between gravity and non-Abelian gauge theories manifests itself in a different set of diagrams that drive the Renormalization Group (RG) flow of the interactions between fermions~\cite{Eichhorn:2011pc}, further analyzed in the broader context of the weak-gravity bound \cite{Eichhorn:2016esv,Christiansen:2017gtg} in asymptotic safety in~\cite{Eichhorn:2017eht}. Additionally, in~\cite{Eichhorn:2016vvy} it has been found that explicit chiral-symmetry breaking operators may be consistently set to zero, in that the gravitational RG flow does not generate them.

These results refer to quantum fluctuations of the metric at fixed topology. More recently, it has been suggested that topological fluctuations, if present in asymptotically safe quantum gravity, might lead to spontaneous chiral-symmetry-breaking~\cite{Hamada:2020mug}.  

Despite these results for quantum fluctuations of the curvature, a classical non-vanishing curvature can act as a source of chiral symmetry breaking. This mechanism is dubbed gravitational catalysis, see, e.g., \cite{Buchbinder:1989fz,Buchbinder:1989ah,Inagaki:1993ya,Sachs:1993ss,Gies:2013dca}. By combining the effect of fluctuations with a classical background, it has been suggested to result in an upper bound on the number of light fermions that can be accommodated in asymptotically safe gravity \cite{Gies:2018jnv}.

Here, we will go beyond a setting with  gravity and fermions and take into account the interplay with quantum fluctuations of an Abelian gauge field. Our motivation lies in tentative hints for a nontrivial asymptotically safe fixed point, that could be induced in the Abelian gauge coupling by quantum gravitational fluctuations \cite{Harst:2011zx,Christiansen:2017gtg,Eichhorn:2017lry}, see also \cite{Daum:2009dn,Folkerts:2011jz,Christiansen:2017cxa}.  Interestingly, in the case without gravity, chiral-symmetry breaking interactions might be present in a nontrivial ultraviolet completion of an Abelian gauge theory \cite{Gies:2020xuh}. Here, we focus on the converse question, namely, whether an ultraviolet completion with a nonzero value of the gauge coupling is made possible without breaking chiral symmetry under the impact of quantum gravitational fluctuations.
As a large enough value of the gauge interaction results in chiral symmetry breaking \cite{Miransky:1984ef,Roberts:1994dr,Aoki:1996fh,Alkofer:2000wg,Kim:2000rr,Gies:2004hy}, a competition between gravitational and effects of the gauge field can be expected. Such an interplay could result in a lower bound on the number of fermions. This expectation arises, since \cite{Harst:2011zx,Christiansen:2017gtg,Eichhorn:2017lry} indicate a decreasing fixed-point value for the gauge coupling as a function of the fermion number.  As the underlying reason for the existence of three generations in the Standard Model is unknown, it is fascinating to understand whether the number of generations is tied to the lightness of fermions, see \cite{Gies:2018jnv}. 

\section{Four-fermion interactions and chiral symmetry breaking}
Let us firstly review the mechanism of chiral symmetry breaking in the presence of an external field, and how it is detected via the RG flow of the system, see, e.g.,  \cite{Gies:2001nw,Gies:2002hq,Gies:2003dp,Braun:2005uj,Braun:2006jd,Braun:2006zz,Braun:2008pi,Haas:2010bw,Braun:2010vd,Braun:2011pp,Braun:2011fw}. To that end, we investigate the fate of chiral symmetry in a fermionic system with chiral $\rm{SU(N_f)_L}\, \times\,\,SU(N_f)_R$ symmetry. Specifically, we study the scale dependence of the two four-fermion operators given by 
\begin{align}
\lambda_+ (V+A)\,,\qquad \lambda_-(V-A)\,,
\end{align}
with
\begin{align}
V=&\left(\bar{\psi}^i\gamma_{\mu}\psi^i\right)\left(\bar{\psi}^j\gamma^{\mu}\psi^j\right)\,,\\
A=&-\left(\bar{\psi}^i\gamma_{\mu}\gamma_5\psi^i\right)\left(\bar{\psi}^j\gamma^{\mu}\gamma_5\psi^j\right)\,,
\end{align}
where the summation over flavor indices runs over $i\in[1,N_{\mathrm{F}}]$.
In the pointlike limit, i.e., without derivatives, these two operators are the only two  Fierz- independent, chirally symmetric four-fermion operators which are Lorentz scalars. 

We will now review the connection between the four-fermion interactions $\lambda_{\pm}$ and the generation of non- trivial fermionic bound states. For this purpose, we first perform a Fierz transformation and change to a scalar-pseudoscalar basis. We focus on $\lambda_{+}$, for which the Fierz transformation reads
\begin{equation}
\lambda_{+}\left(V+A\right)=\lambda_{\sigma}\left[(\bar{\psi}^i\psi^j)^2-(\bar{\psi}^i\gamma_5\psi^j)^2\right]\,,
\end{equation}
with $(\bar{\psi}^i\psi^j)^2=(\bar{\psi}^i\psi^j)(\bar{\psi}^i\psi^j)$ and similarly for the pseudoscalar channel. This relation is an exact Fierz identity if \cite{Gies:2001nw,Braun:2011pp}
\begin{equation}
\lambda_{\sigma}=-\frac{1}{2}\lambda_{+}\,.
\end{equation}

Using a Hubbard-Stratonovich transformation, see, e.g., \cite{Klevansky:1992qe,Buballa:2003qv,Braun:2011pp}, the four-fermion interactions can be rewritten in terms of auxiliary fields. We will now focus on the case of one single flavor, but the following arguments generalize to multiple flavors. For the scalar part of the interaction, the four-fermion interactions can be rewritten as
\begin{equation}
-\frac{\lambda_{\psi}}{4}(\bar{\psi}\psi)^2=\left[h(\bar{\psi}\psi)\sigma+m_{\varphi}^2\sigma^2\right]_{\mathrm{EoM}(\sigma)}\,,
\end{equation}
with 
\begin{equation}
\label{eq: HubbStrat}
m_{\varphi}^2=\frac{h^2}{\lambda_{\psi}}\,,
\end{equation}
which holds on the equations of motion for the scalar field $\sigma$, i.e., when the auxiliary field is integrated out. In terms of the scalar degree of freedom, $\sigma$, 
chiral symmetry is spontaneously broken when the mass term $m_{\varphi}^2$ becomes negative. In terms of the original fermionic degrees of freedom, the onset of chiral symmetry breaking is therefore indicated by a divergence of the four-fermion interaction $\lambda_{\psi}$, cf.~Eq.~\eqref{eq: HubbStrat}. 
This argument, exemplified for the scalar channel, applies to the other channels in a similar way. Since the breaking of chiral symmetry could in principle remain isolated in one single channel, it is important to consider a complete basis of four-fermion interactions. 

Furthermore, explicit studies in QCD show that the RG-scale $k_{\mathrm{\chi\mathrm{SB}}}$, where the four-fermion interaction diverges, approximately corresponds to the physical scale of chiral symmetry breaking \cite{Ellwanger:1994wy,Jaeckel:2003uz,Braun:2005uj}. This is a crucial motivation for studies in a gravitational context, where $k_{\mathrm{\chi\mathrm{SB}}}$ would be expected to lie in the vicinity of the Planck scale, resulting in Planckian bound-state masses.

In summary, the question of spontaneous breaking of chiral symmetry can be studied by investigating the fixed-point structure of the four-fermion interactions $\lambda_{\pm}$. Schematically, within the FRG, their scale dependence is given by
\begin{equation}
\label{eq: betaschem}
k\partial_k \lambda_{\pm}=2\lambda_{\pm}+\sum_{i=0}^{2}a_i^{\pm}\,\lambda_{+}^i\,\lambda_{-}^{2-i}+b_{\pm}\lambda_{\pm}\,h_{\mathrm{ext}}+c_{\pm}h_{\mathrm{ext}}^2\,,
\end{equation} 
where $a_i^{\pm}$, $b_{\pm}$ and $c_{\pm}$ are numerical coefficients, and $k$ is the RG scale. 
Furthermore, $h_{\mathrm{ext}}$ is an external coupling associated with vertices encoding the interactions of  fermions with other fields, such as a gauge field or gravity.

\begin{figure}[t!]
	\centering
	\includegraphics[width=.9\linewidth]{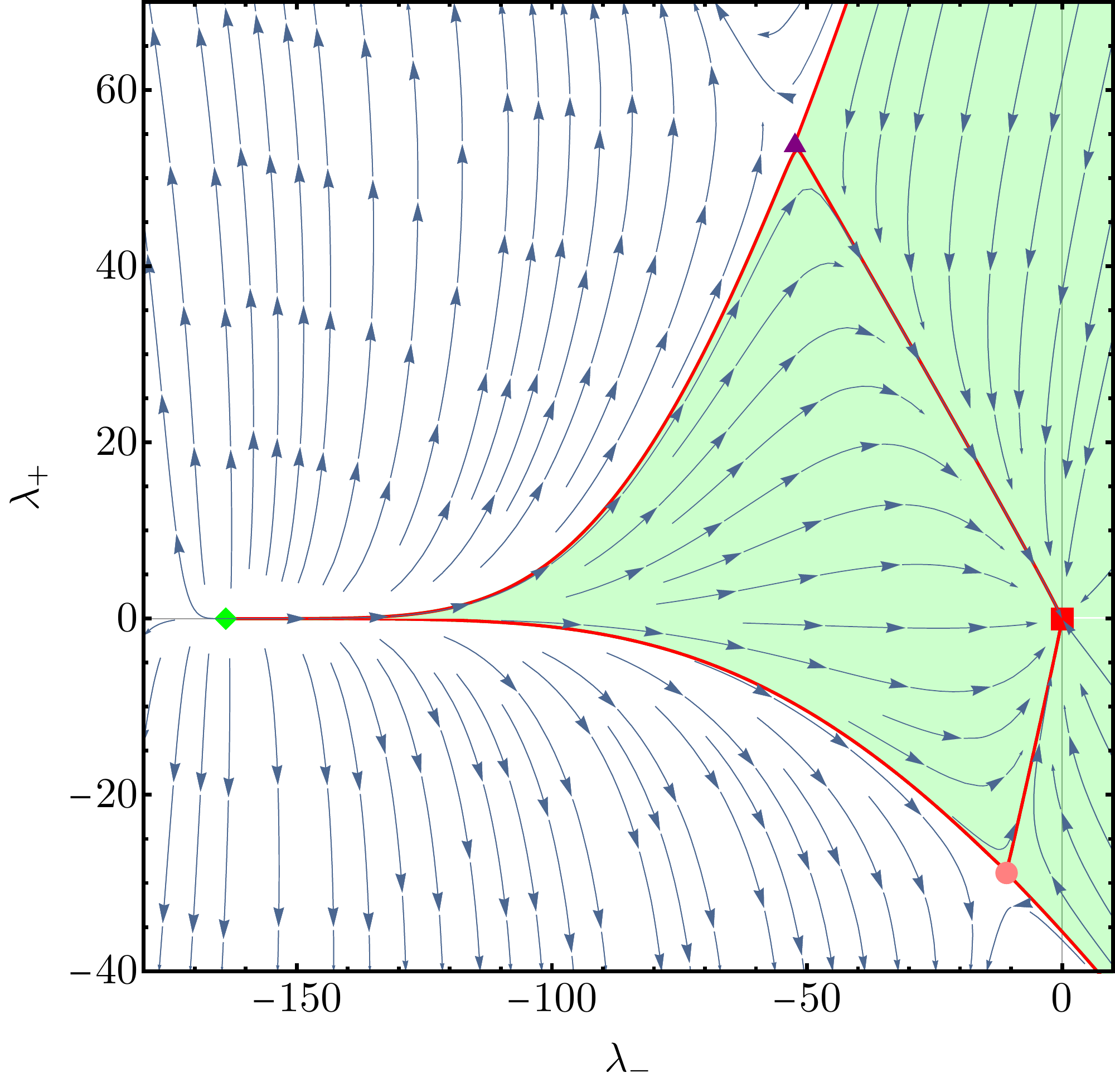}
	\caption{The green region marks the region of initial conditions $\lambda_{\pm,\mathrm{in}}$, for which spontaneous chiral symmetry breaking can be avoided. For initial conditions in the white regions,  one of the four-fermion couplings will be driven towards criticality, signaling the onset of spontaneous chiral symmetry breaking. We choose $N_{\mathrm{F}}=2$, $G=\Lambda=0$, and $e=1$ for this illustration.}
	\label{fig: Strem4FPPlot}
\end{figure}
\begin{figure}
	\centering
	\includegraphics{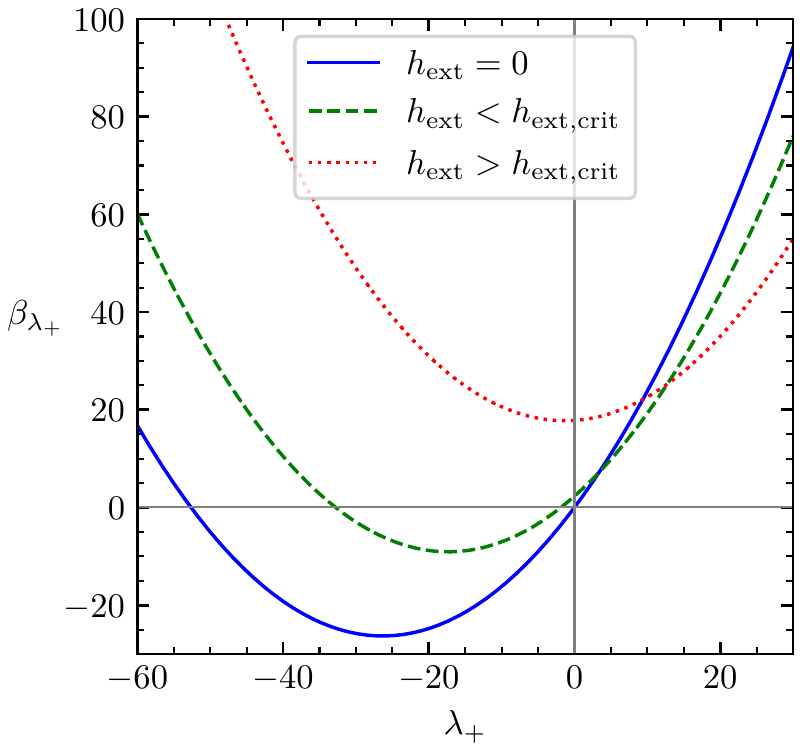}
	\caption{We illustrate the scale dependence of one of the four-fermion interactions $\lambda_+$, for the choice of $\lambda_-=0$ under the impact of an external gauge field $h_{\mathrm{ext}}$ and in the absence of gravity. The solid blue line illustrates the case where the external gauge field is absent. In this case, $\beta_{\lambda_+}$ features a free, infrared (IR) attractive fixed point, as well as an interacting, IR repulsive fixed point. Increasing the strength of the external gauge coupling $e$, both fixed points approach each other (green, dashed line), until they collide and annihilate for a critical value $h_{\mathrm{ext},\mathrm{crit}}$. This annihilation is an indicator of chiral-symmetry breaking, as observed in QCD \cite{Gies:2002hq,Braun:2005uj,Gies:2005as,Braun:2006jd}. 
	}
	\label{fig: betaplots}
\end{figure}

In the absence of an additional field, i.e., for $h_{\mathrm{ext}}=0$, the combined set of $\beta$-functions has four fixed points, one of which is Gaussian, $\lambda_{\pm,\, \ast}=0$, while the other ones are interacting. Due to the higher-order character of the fermionic self-interaction operators, the four-fermion interactions are irrelevant, i.e., IR attractive, at the Gaussian fixed point, where $\lambda_{+,\, *}=\lambda_{-,\,*}=0$. In contrast, they have at least on IR repulsive, i.e., relevant, direction at each of the three interacting fixed points.
Thus, specifying to the case illustrated in Fig.~\ref{fig: Strem4FPPlot}, all initial conditions for the RG flow lying outside the green region 
lead to spontaneous breaking of chiral symmetry at  a scale $k_{\mathrm{\mathrm{\chi\mathrm{SB}}}}>0$, where one of the couplings $\lambda_{\pm}(k)$ diverges.

A qualitatively similar fixed-point structure persists for sufficiently small external coupling $|h_{\mathrm{ext}}|<|h_{\mathrm{ext},\,\mathrm{crit.}}|$.  The fixed-point structure is then still dominated by the first two terms in Eq.~\eqref{eq: betaschem}, while non-vanishing fixed-point values for $\lambda_\pm$ are now unavoidable, due to the last, $\lambda_\pm$-independent term. In particular, the Gaussian fixed point at $\lambda_{\pm,\, \ast}=0$ gets shifted to a finite fixed-point value, cf.~Fig.~\ref{fig: betaplots}.
Depending on the sign of the coefficient $c$ in Eq.~\eqref{eq: betaschem}, there is a critical value $h_{\mathrm{ext},\,\mathrm{crit.}}$, such that at  $|h_{\mathrm{ext}}|=|h_{\mathrm{ext},\,\mathrm{crit.}}|$, the two fixed points collide and annihilate. Thus,
for $|h_{\mathrm{ext}}|>|h_{\mathrm{ext},\,\mathrm{crit.}}|$ the zeros of $\beta_{\lambda_\pm}$ lie off the real axis, as illustrated in Fig.~\ref{fig: betaplots}. 
In this case, any initial condition for $\lambda_{\pm}$ will lead to spontaneous breaking of chiral symmetry. Therefore, large external couplings $h_{\mathrm{ext}}$ can drive a chirally symmetric system to criticality.  Indeed, this mechanism determines the IR spectrum of bound states in QCD, where the non-Abelian gauge coupling becomes strong in the IR, inducing a divergence in the four-fermion interactions, see, e.g., \cite{Alkofer:2000wg,Gies:2002hq,Braun:2005uj,Gies:2005as,Braun:2006jd,Braun:2006zz,Braun:2008pi,Haas:2010bw,Braun:2010vd,Braun:2011pp,Braun:2011fw}.
Similar mechanisms occur in QED in three and four dimensions, where the Abelian gauge coupling $e$ becomes strong in the UV \cite{Appelquist:1986fd,Gockeler:1997dn,Braun:2014wja}. 

While the parameter $c$ for gravity in Eq.~\eqref{eq: betaschem} 
has the appropriate sign to lead to a collision of fixed points, the gravitational contribution to $b_{\pm}$ compensates this effect. Overall, in the absence of an Abelian gauge field, quantum gravitational fluctuations stabilize the system, since the fixed points are actually driven away from a collision with an increasing strength of the gravitational interactions \cite{Eichhorn:2011pc}. 

We now add the Abelian gauge field to the setting.
In the context of asymptotically safe quantum gravity, there are indications for a gravity-generated interacting fixed point of the Abelian gauge coupling $e$ \cite{Harst:2011zx,Christiansen:2017gtg,Eichhorn:2017lry}. For gravity itself, compelling evidence for a gravitational UV fixed point, also known as the Reuter fixed point \footnote{For introductions and reviews, see, e.g., \cite{Niedermaier:2006ns,Percacci:2007sz,Reuter:2012id,Eichhorn:2017egq,Eichhorn:2018yfc,Pereira:2019dbn,Reichert:2020mja,Eichhorn:2020mte,Pawlowski:2020qer}, and \cite{Bonanno:2020bil} for a critical discussion of the current status of asymptotically safe gravity.}, has been found in various approximations to the gravitational dynamics in pure gravity \cite{Souma:1999at,Reuter:2001ag, Lauscher:2001ya,Litim:2003vp,Machado:2007ea,Codello:2008vh, Benedetti:2009rx,Machado:2009ph,Manrique:2009uh,Manrique:2010am,Groh:2010ta,Eichhorn:2010tb, Manrique:2011jc,Benedetti:2012dx,Rechenberger:2012dt,Christiansen:2012rx,Dietz:2012ic,Ohta:2013uca,Falls:2013bv,Falls:2014tra,Christiansen:2014raa,Becker:2014qya,Christiansen:2015rva,Morris:2015oca,Ohta:2015efa,Ohta:2015fcu,Gies:2015tca, Demmel:2015oqa, Biemans:2016rvp,Gies:2016con, Denz:2016qks,Platania:2017djo,Falls:2017lst, Knorr:2017fus,Christiansen:2017bsy,deBrito:2018jxt,Falls:2018ylp,Kluth:2020bdv,Falls:2020qhj} and gravity-matter systems  \cite{Narain:2009fy,Shaposhnikov:2009pv,Dona:2013qba,Meibohm:2015twa,Dona:2015tnf,Oda:2015sma,Eichhorn:2016vvy,Wetterich:2016uxm,Biemans:2017zca,Christiansen:2017cxa,Hamada:2017rvn,Eichhorn:2017als,Eichhorn:2017ylw,Eichhorn:2017sok,Alkofer:2018fxj,Eichhorn:2018akn,Eichhorn:2018ydy,Eichhorn:2018nda,Pawlowski:2018ixd,Knorr:2019atm,Burger:2019upn,Eichhorn:2019yzm,Reichert:2019car,Daas:2020dyo,Eichhorn:2020kca,Eichhorn:2020sbo}. In particular, the impact of gravity on the interaction structure of fermionic systems and gauge fields has been explored in \cite{Zanusso:2009bs,Vacca:2010mj,Harst:2011zx,Eichhorn:2011pc,Oda:2015sma,Eichhorn:2016esv,Christiansen:2017gtg,Hamada:2017rvn,Christiansen:2017qca,Eichhorn:2017lry,Eichhorn:2017ylw,Eichhorn:2018whv,deBrito:2019umw}.
For gauge fields, a universal, i.e., gauge-group independent  gravitational contribution $f_g$ to the scale-dependence of $e$ arises. Within our truncation, it reads
\begin{equation}
\label{eq: deshem}
k\partial_k e=-f_g\, e+\frac{N_{\mathrm{F}}}{24 \pi^2}\,e^3\,,
\end{equation}
where $f_g\geq0$ was found within FRG studies \cite{Daum:2009dn,Harst:2011zx,Folkerts:2011jz,Christiansen:2017gtg,Eichhorn:2017lry,Christiansen:2017cxa,Eichhorn:2019yzm,deBrito:2019umw}. At the interacting fixed point $e_{*,\mathrm{int}}$, the Abelian gauge coupling corresponds to an irrelevant direction in theory space. Thus, such a fixed point would not only constitute a potential solution to the triviality problem, but, more importantly, provide a first-principles derivation of the IR-value of the gauge coupling \cite{Harst:2011zx,Eichhorn:2017lry}.

The size of the fixed point is actually determined by the gravitational contribution, $f_g$, as well as the number of fermions, $N_F$, i.e., 
\begin{eqnarray}
e_{\ast,\, \rm int}= \sqrt{\frac{24\pi^2\, f_g}{N_{\rm F}}}.\label{eq:estar}
\end{eqnarray}
Depending on the number of fermions and the gravitational interactions, the fixed-point value for the Abelian gauge coupling $e$ might accordingly lie in a strong-coupling regime and could potentially drive the fermions to criticality. The exact value of the interacting fixed point for $e$ in Eq.~\eqref{eq:estar} actually also depends on $N_{\rm F}$ implicitly, through the $N_{\rm F}$ dependence of $f_g$. In this paper, we will accordingly investigate, whether there is a critical number of fermions, for which the interacting fixed point for $e$ drives the fermionic interactions to criticality. 
\section{Functional Renormalization Group setup}
To extract the scale dependence of the two four-fermion couplings $\lambda_{\pm}$, we employ the FRG \cite{Wetterich:1992yh,Ellwanger:1993mw,Morris:1993qb,Reuter:1993kw}.  It implements the Wilsonian idea of integrating out quantum fluctuations according to their momentum shell,  based on a functional integro-differential equation.  The Wetterich equation \cite{Wetterich:1992yh} allows to extract the scale dependence of couplings within and beyond perturbation theory and is given by
\begin{equation}\label{eq: WEQ}
k\partial_k \Gamma_{k}=\frac{1}{2}\mathrm{STr}\left(k\partial_k R_k \left(\Gamma_k^{(2)}+R_k\right)^{-1}\right)\,.
\end{equation}
Here, $\Gamma_{k}$ is the scale dependent effective action, interpolating between the microscopic action $\Gamma_{k\rightarrow \infty}$ and the full quantum effective action $\Gamma_{k=0}$. Eq.~\eqref{eq: WEQ} gives rise to flow equations for the couplings of the system, which encode how the effective dynamics changes, as quantum fluctuations with momenta of the order $k$ are integrated out. For more details on the FRG, as well as applications to gauge theories and gravity, see, e.g., \cite{Berges:2000ew,Pawlowski:2005xe,Gies:2006wv,Rosten:2010vm,Dupuis:2020fhh}. 

For our practical studies, we will truncate the  scale dependent effective action $\Gamma_k$ to a finite set of interactions, thus introducing a systematic uncertainties into the beta functions and fixed points of interest. Our truncation follows the physical principle of a near-perturbative asymptotically safe fixed point. Such a fixed point is characterized by near- canonical scaling, closely following the canonical power counting of operators.  The expectation of a near-perturbative nature of the Reuter fixed point is supported by various studies, including indications for the near-perturbative nature of quantum gravity via symmetry identities in gravity-matter systems \cite{Denz:2016qks,Eichhorn:2018nda,Eichhorn:2018ydy,Eichhorn:2018akn}, the study of near-Gaussian scaling in extended pure-gravity  \cite{Falls:2013bv,Falls:2014tra,Falls:2017lst,Falls:2018ylp,Kluth:2020bdv} and critical exponents in gravity-matter truncations \cite{Narain:2009fy,Narain:2009gb,Eichhorn:2011pc, Percacci:2015wwa,Eichhorn:2016esv,Eichhorn:2016vvy,Eichhorn:2017eht,Eichhorn:2017sok,Eichhorn:2018nda,Daas:2020dyo,Eichhorn:2020sbo}, as well as indications from one-loop perturbation theory \cite{Niedermaier:2009zz,Niedermaier:2010zz}.

For the present investigation, we approximate the dynamics of the fermionic subsystem by
\begin{align}
\label{eq: ferm}
&\Gamma_{k,\,\mathrm{F}}=\int\!\!\mathrm{d}^4 x \sqrt{g}\, i\, Z_{\psi}(k)\, \bar{\psi}^i\gamma^{\mu}\nabla_{\mu}\psi^i\\
&+\frac{Z_{\psi}(k)^2}{2k^2}\int \!\!\mathrm{d}^4 x \sqrt{g}\,\left[\lambda_{-}(k)\left(V-A\right)+\lambda_{+}(k)\left(V+A\right)\right]\,,\notag
\end{align}
where $Z_{\psi}$ is the fermion wave-function renormalization, giving rise to the anomalous dimension
\begin{equation}
\eta_{\psi}=-k\partial_k \ln Z_{\psi}\,.
\end{equation}

The minimal coupling of the fermions to the Abelian gauge field, as well as to gravity is encoded in the covariant derivative in the fermionic action \eqref{eq: ferm}, which is given by
\begin{equation}\label{eq: covder}
	\nabla_{\mu}=\partial_{\mu}+i e A_{\mu}+\frac{1}{8}[\gamma^a,\gamma^b]\omega_{\mu}^{ab}\,,
\end{equation}
with Abelian gauge coupling $e$ and the spin connection $\omega^{\mu}_{ab}$. Within the formalism considered here, the spin-connection is not treated as an independent variable. In this case, $\omega^{\mu}_{ab}$ can be determined in terms of the Christoffel connection in the Vielbein formalism. Equivalently, the covariant derivative can be expressed in the spin-base invariance formalism \cite{Gies:2013noa,Gies:2015cka,Lippoldt:2015cea}. Studies of non-minimally coupled fermion-gravity systems have been put forward in \cite{Eichhorn:2016vvy,Eichhorn:2018nda,Daas:2020dyo}; and regulator studies in the context of gravity-fermion systems can be found in \cite{Dona:2012am}.

For the dynamics of the Abelian gauge field $A_{\mu}$, we consider the truncation
\begin{equation}
\Gamma_{k,\,U(1)}=\frac{ Z_A(k)}{4}\int\!\!\mathrm{d}^4 x \sqrt{g}\,g^{\rho\mu}g^{\kappa\nu}F_{\mu\nu}F_{\rho\kappa}\,,
\end{equation}
where $F_{\mu\nu}$ is the field strength tensor and  $Z_A$ is the gauge-field wave-function renormalization, giving rise to the anomalous dimension for the gauge field
\begin{equation}
	\eta_{A}=-k\partial_k \ln Z_A\,.
\end{equation}
The beta function for the Abelian gauge coupling $e(k)$ can be computed in terms of the anomalous dimension via
\begin{align}
\label{eq: WID}
\beta_e = \frac{e}{2} \, \eta_A .
\end{align}

In the gravitational sector we consider the Einstein-Hilbert approximation, i.e.,
\begin{equation}
\Gamma_{k,\,\mathrm{Grav}}= \frac{k^2}{16\pi G(k)}\int\!\!\mathrm{d}^4 x \sqrt{g}\left(2k^2\Lambda(k) - R\right)\,,
\end{equation}
where $R$ is the Ricci-curvature, and $\Lambda(k)$ and $G(k)$ are the dimensionless counterparts of the scale-dependent cosmological constant and the Newton coupling, respectively. 

Employing a local coarse-graining procedure like the FRG requires the introduction of a background metric $\bar{g}_{\mu\nu}$. It has the role of a non-dynamical field and -- in principle -- does not need to be specified. However, specific choices of the background metric $\bar{g}_{\mu\nu}$ greatly simplify computations. In the present case, where we are mostly interested in the scale dependence of curvature independent, pure matter couplings, a flat background, i.e.,
\begin{equation}
\bar{g}_{\mu\nu}=\delta_{\mu\nu}\,,
\end{equation}
is the simplest choice. The Ansatz for the scale-dependent effective action $\Gamma_k$ is expanded in terms of metric fluctuations
\begin{equation}
h_{\mu\nu}=g_{\mu\nu}-\delta_{\mu\nu}\,,
\end{equation}
where $h$ does not need to be small despite the suggestive notation.

To complete our truncation for $\Gamma_k$, we add the standard gauge-fixing for the Abelian gauge field 
\begin{equation}
\Gamma_{k, U(1), \mathrm{gf}}=\frac{1}{\zeta}\int\!\!\mathrm{d}^4x \sqrt{\bar{g}}\, (\bar{g}^{\mu\nu}\bar{D}_{\mu}A_\nu)^2\,,\qquad \zeta\to0\,,
\end{equation}
and for gravity
\begin{equation}
\Gamma_{k, \mathrm{grav}, \mathrm{gf}}=\frac{1}{\alpha}\frac{1}{32\pi G}\int\!\!\mathrm{d}^4x \sqrt{\bar{g}}\mathcal{F}^{\mu}\bar{g}_{\mu\nu}\,\mathcal{F}^{\nu}\,,\quad \alpha\to 0\,,
\end{equation}
with the gauge-fixing condition 
\begin{equation}
\mathcal{F}^{\mu}=\left(\bar{g}^{\mu\kappa}\bar{D}^{\lambda}-\frac{1+\beta}{4}\bar{g}^{\kappa\lambda}\bar{D}^{\mu}\right)h_{\kappa\lambda}\,,\qquad \beta\to\alpha\to0\,.
\end{equation}
Here, $\bar{D}^{\mu}$ represents the space-time covariant derivative defined with respect to the background metric $\bar{g}_{\mu\nu}$. The corresponding Faddeev-Popov ghost terms contribute to the gravitational beta functions.

\begin{figure}
	\centering
	\includegraphics[width=\linewidth]{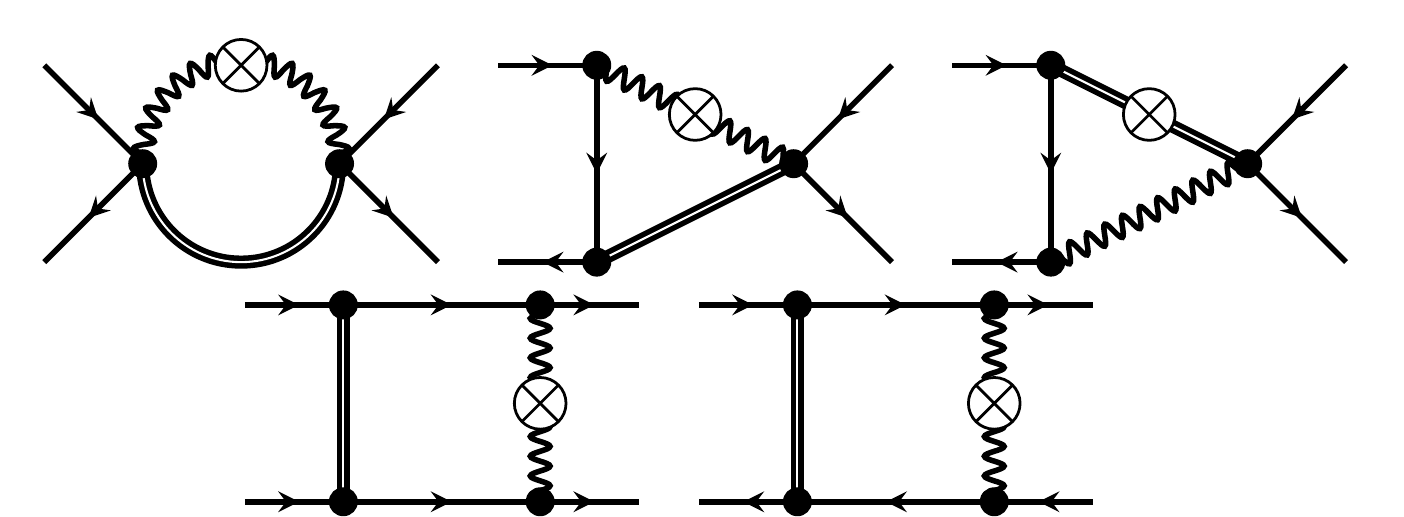}
	\caption{Diagrams containing internal vector bosons (curly lines), metric fluctuations (double lines) and fermions (solid lines), contributing to the scale dependence of four fermion interactions $\lambda_{\pm}$. The regulator insertion, marked as a cross, has to be inserted on each one of the internal lines, such that various regularized versions exist for each diagram displayed here.}
	\label{fig: diagrs}
\end{figure}

The flow equation \eqref{eq: WEQ} resulting from this setup can be rewritten in terms of one-loop-type (nonperturbative) diagrams.
The diagrams contributing to the scale dependence of $\lambda_{\pm}$ can be classified according to the internal lines. For the diagrams with internal fermion and gauge-field lines, we refer to \cite{Gies:2003dp}, while for internal fermions and metric-fluctuations, we refer to \cite{Eichhorn:2011pc}. Our results agree with those provided in these references. The remaining diagrams contributing to $\beta_{\lambda_{\pm}}$ contain internal vector bosons and metric fluctuations, cf.~Fig.~\ref{fig: diagrs}. 

The regulator $R_k$ in Eq.~\ref{eq: WEQ} inherits its tensor structure from the two-point vertex $\Gamma_k^{(2)}$ \cite{Pawlowski:2001df,Gies:2002af,Pawlowski:2005xe,Gies:2015tca}, resulting in a form which is diagonal in field space, with components given by
\begin{equation}
	R_k(p^2)=\Gamma_k^{(2)}(p^2)\,r_k(p^2/k^2)\big|_{\Lambda=\lambda_+=\lambda_-=0}\,.
\end{equation}
Here, the evaluation of $\Gamma_k^{(2)}$ at $\Lambda=\lambda_+=\lambda_-=0$  ensures that no momentum independent contributions enter the regulator $R_k(p^2)$ \cite{Meibohm:2015twa}, and where the two derivatives in $\Gamma^{(2)}_k(p^2)$ refer to two derivatives with respect to the  fields. 
 For the shape functions, we choose Litim-type cutoffs \cite{Litim:2001up}, i.e.,
\begin{align}
r_k^{h}(x)=r_k^{A}(x)=&\left(\frac{1}{x}-1\right)\theta\left(1-x\right)\\
r_k^{\psi}(x)=&\left(\frac{1}{\sqrt{x}}-1\right)\theta\left(1-x\right)\,.
\end{align}
Since we explore a potential mechanism of chiral-symmetry-breaking, it is important to choose regulators that respect the chiral symmetry. Explicitly, the regulator for the fermions is given by
\begin{equation}
	R_k(p^2)=Z_{\psi}\slashed{p}\left(\sqrt{\frac{k^2}{p^2}}-1\right)\theta\left(1-\frac{p^2}{k^2}\right)\,,
\end{equation}
and the corresponding generalization in terms of $\slashed{\nabla}$ for the fermionic contribution to the gravitational beta functions.

We evaluate the scale dependence in a perturbative approximation, where the anomalous dimensions coming from the regulator insertion, $k\partial_k R_k$, are neglected. To evaluate the RG flow, we use the Mathematica packages \emph{xAct} \cite{Brizuela:2008ra,Martin-Garcia:2007bqa,MartinGarcia:2008qz,2008CoPhC.179..597M} and \emph{DoFun}~\cite{Huber:2011qr,Huber:2019dkb}, as well as the \emph{FormTracer}~\cite{Cyrol:2016zqb}.

\section{Results: Light fermions in asymptotic safety and beyond}
In our approximation,
the scale dependence of the four-fermion interactions $\lambda_{\pm}$ reads:
\begin{align}\label{eq: betaplusminus}
\beta_{\lambda_{\pm}}&=\,\,2\lambda_{\pm}+M_{\pm} 
-\frac{5 \lambda _{\pm} G}{8 \pi  (1-2 \Lambda )^2}\pm\frac{5 G^2}{8 (1-2 \Lambda )^3} \\
&+\frac{3 \lambda _{\pm} G}{4 \pi  (3-4 \Lambda )}+\frac{15 \lambda _{\pm} G}{8 \pi  (3-4 \Lambda )^2}
+\frac{5 e^2 G}{16 \pi  (1-2 \Lambda )} \notag\\
&+\frac{5 e^2 G}{16 \pi  (1-2 \Lambda )^2} +\frac{9 e^2 G}{160 \pi  (3-4 \Lambda )}-\frac{27 e^2 G}{160 \pi  (3-4 \Lambda )^2}\,,\notag
\end{align}
where the matter contributions $M_{\pm}$ are given by
\begin{align}
M_+=\,\,&\frac{8 \lambda _+ \left(\lambda _- \left(N_F+1\right)-3 e^2\right)+9 e^4+12 \lambda _+^2}{32 \pi ^2}\,,\\
M_-=\,\,&\frac{4 \lambda _-^2 \left(N_F-1\right)+4 \lambda _+^2 N_F+24 \lambda _- e^2-9 e^4}{32 \pi ^2}\,.
\end{align}
The first term in Eq.~\eqref{eq: betaplusminus} is due to the canonical mass dimension of the operators, while the matter contribution $M_{\pm}$ arises from fermionic and gauge field contributions, which can also be found in \cite{Gies:2003dp}. From the third to the sixth term we have the gravitational contributions to the scale dependence of $\lambda_{\pm}$, cf.~\cite{Eichhorn:2011pc}. Finally, terms involving $e^2 G$ represent the contributions of the diagrams shown in Fig.~\ref{fig: diagrs}.

This system features four zeros of the beta function; the most predictive one, the shifted Gaussian fixed point, features two irrelevant directions and will be our main focus in the following.

In the current setup, the beta function for the gauge coupling is given by \cite{Harst:2011zx,Christiansen:2017gtg,Eichhorn:2017lry,Eichhorn:2019yzm}:
\begin{align}
	\beta_e=&\left(-\frac {5 G} {9\pi (1-2\Lambda)} + \frac {5 G} {18\pi (1 - 2\Lambda)^2}\right)\,e+ \frac{e^3 N_{\mathrm F}}{24 \pi ^2}\,. \label{eq:betae}
\end{align} 

 \subsection{Asymptotic safety and chiral symmetry}

In order to explore the asymptotically safe case, we supplement the beta functions in the matter sector by those in the gravitational one. We approximate the scale dependence of $G$ and $\Lambda$ by the results obtained in \cite{Dona:2012am,Eichhorn:2016vvy}, adding the contribution from the Abelian gauge field from \cite{Dona:2013qba}. Specifically, the beta-functions for the gravitational couplings are given by
\begin{eqnarray}
\label{eq: betaglBG}
\beta_{G}&= &\,2 G-G^2\, \eta_g\,,\\
\beta_{\Lambda}&=&-2\Lambda-G\Lambda\, \eta_g \label{eq:betalambda}\\
&{}&-\frac{G}{2 \pi }\notag  \left(\frac{5}{4 \Lambda -2}+\frac{3}{8 \Lambda -6}+2 N_{\text{F}}+6-8 \log \left(\frac{3}{2}\right)\right)
\,,
\end{eqnarray}
with
\begin{align}
\eta_g=\frac{1}{12 \pi } \bigg( &\frac{6}{4 \Lambda -3}+\frac{10}{1-2 \Lambda }+\frac{20}{(1-2 \Lambda )^2} \nonumber\\
&-4 N_{\text{F}}+19+32 \log \left(\frac{3}{2}\right) \bigg)\,.
\end{align}
These are obtained within a truncation accounting for the effect of minimally coupled matter fields on the gravitational couplings.
In the absence of fermionic self-interactions, the inclusion of induced non-minimal interactions does not lead to a qualitative change of the fixed-point structure, cf.~\cite{Eichhorn:2018nda}.
This set of beta functions admits a fixed point at finite fermion numbers; as in \cite{Dona:2013qba,Eichhorn:2016vvy}, the fixed point cannot be extended beyond $N_{\rm F} =7$. This could change under the extension to a fluctuation setting \cite{Meibohm:2015twa,Eichhorn:2018nda,Pawlowski:2020qer}. As our focus is on small fermion numbers here, where all these studies agree on the existence of a fixed point, we work with the above beta functions in this paper. We also comment that an effective strength of the gravitational coupling, which is the key quantity to understand the gravitational impact on the matter system, actually features a qualitatively similar dependence on the fermion number for both background and fluctuation studies, see the discussion in \cite{Eichhorn:2018nda}.

\subsubsection{Fixed-point collisions at finite fermion number}

\begin{figure}[t!]
	\centering
	\includegraphics[width=.9\linewidth]{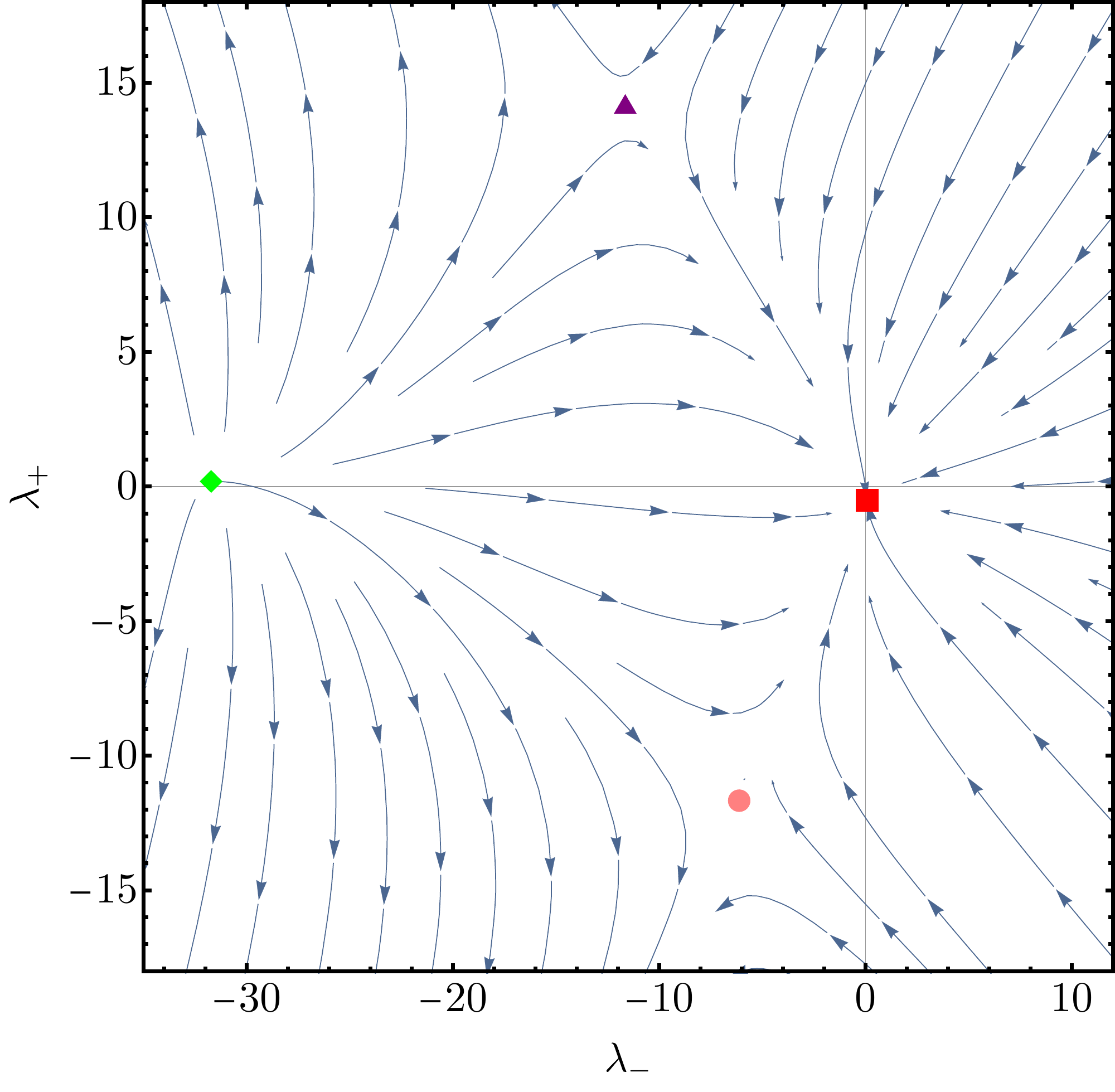}
	\begin{minipage}{.85\linewidth}
	\includegraphics[width=1\linewidth]{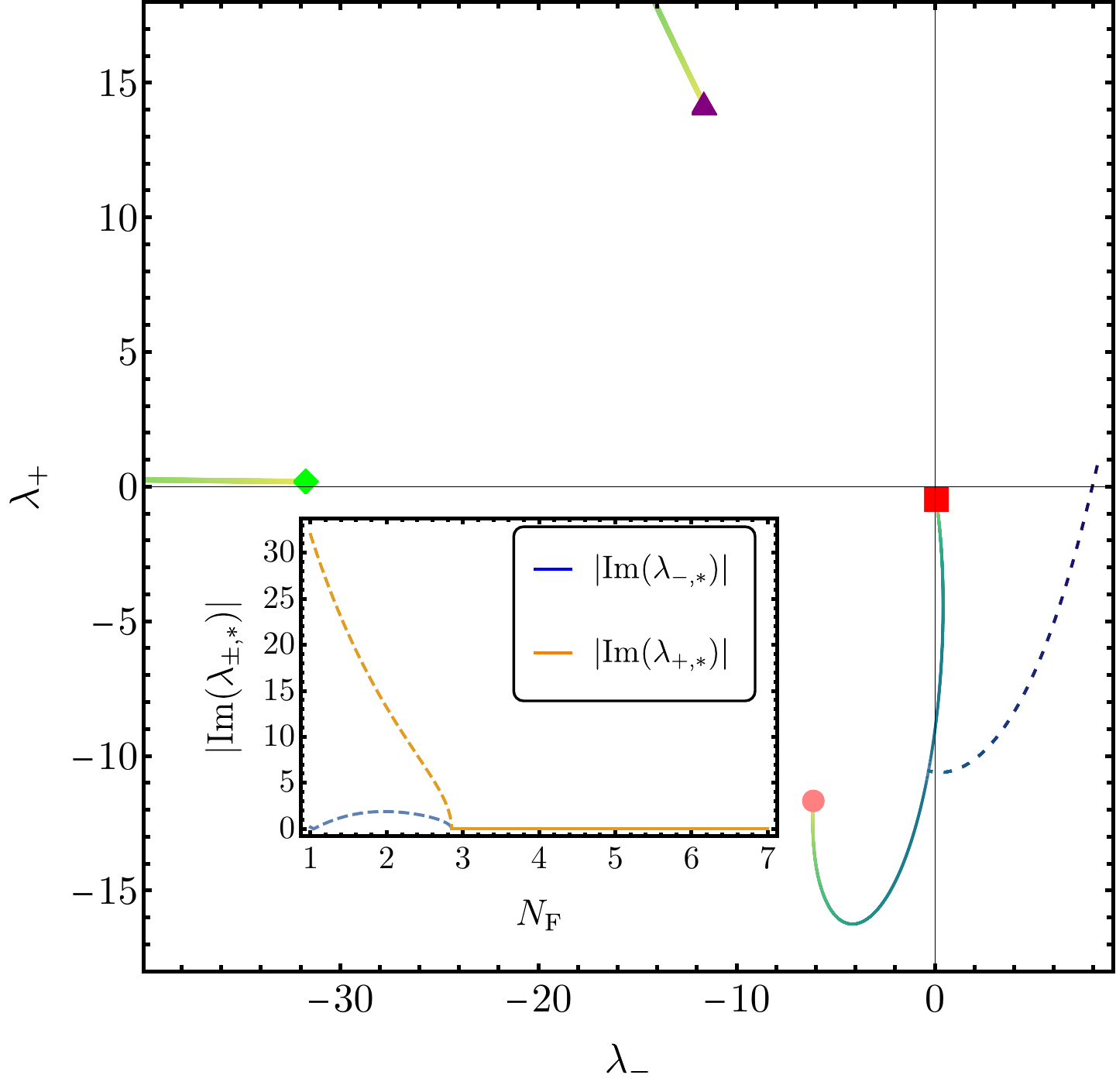}
	\end{minipage}
	\begin{minipage}{.08\linewidth}
		\vspace{-18pt}
	\includegraphics[width=.8\linewidth]{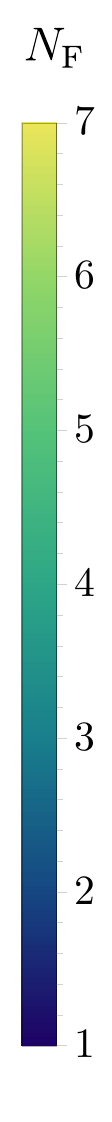}
	\end{minipage}
	\caption{Upper panel: fixed-point-structure of the system spanned by $\lambda_{\pm}$, evaluated on the fixed point values for $G,\Lambda$, and on the interacting fixed point for $e$, at $N_{\mathrm{F}}=7$. The shifted Gaussian fixed point, which is the most predictive fixed point featuring two irrelevant directions is marked as  red box, while two other fixed points (pink circle and purple triangle) both feature one relevant and one irrelevant direction. The last fixed point (green diamond) has two relevant directions.\\
		Lower panel: The evolution of all four fixed points for $\lambda_{\pm}$ is shown as a function of the number $N_{\mathrm{F}}$ of Dirac fermions. While two fixed points move to larger absolute values of $\lambda_{\pm}$, when lowering $N_{\mathrm{F}}$, the two other fixed points collide at $N_{\mathrm{F}}\approx2.9$, where they vanish into the complex plane. After the collision, we show the real part of the fixed point as dashed line, while the inset shows the absolute value of the imaginary parts.}
	\label{fig: FPstruct}
\end{figure}

We  consider the fully dynamical system of beta functions in the matter sector, $\beta_{\lambda_{\pm}}, \beta_e$. At given values of the gravitational couplings $G$ and $\Lambda$, it can feature a total of up to eight phenomenologically viable fixed points, four at $e_{\ast}=0$ and four at  $e_{\ast}>0$. The case with $e_{\ast}=0$ exists for small $N_F$, and has been investigated in \cite{Eichhorn:2011pc}. We instead focus on the most predictive fixed point and explore its fate as a function of $N_F$. Promoting $G$ and $\Lambda$ to running couplings provides us with a set of five beta functions to solve. In our approximation, the interactions in the matter sector do not ``backreact" on the gravitational sector, such that we first determine $G_{\ast},\, \Lambda_{\ast}$ from  Eq.~\eqref{eq: betaglBG} and \eqref{eq:betalambda} and subsequently search for a fixed point in the matter sector. The four fixed points exist at $e_{\ast}>0$ for $2.9\lesssim N_{\mathrm{F}}\leq7$ cf.~Fig.~\ref{fig: FPstruct}, where $e_{\ast}=e_{\ast}(N_{\mathrm{F}})$, cf.~Fig.~\ref{fig: ecrit}
\footnote{The limiting value $N_{\mathrm{F}}\leq7$ arises in the background-field approximation \cite{Dona:2013qba,Eichhorn:2016vvy,Daas:2020dyo}; it has not been discovered in truncations in the fluctuation setting  \cite{Meibohm:2015twa,Eichhorn:2018nda,Pawlowski:2020qer}. Since the present analysis gives rise to a lower bound on the number of fermions well below this value,
its existence is not important for the qualitative result of this section.}. 
In particular, the most predictive fixed point, in which all three matter couplings are irrelevant, and which corresponds to the Gaussian fixed-point for $\lambda_{\pm}$, only appears out of the complex plane upon a fixed-point collision with a less predictive fixed point at $N_F\approx 2.9$, cf. Fig.~\ref{fig: FPstruct}. Therefore, the realization of the interacting fixed point $e_{*}$ does not only predict the IR value of the gauge coupling, but also puts a lower bound on the number of light fermions. Specifically, more than $N_{\mathrm{F}}=3$ Dirac fermions are required in our truncation to allow for the existence of light fermions in the IR.

While the previous discussion was focused on the 
interacting fixed point $e_{*}>0$, a similar lower bound on the number of fermions 
arises when the gauge coupling becomes asymptotically free. In this case, the scale-dependent gauge coupling $e(k)$ increases towards the IR. Its Planck-scale value is a free parameter of the setting, but is limited by 
the maximum value $e_{\mathrm{max}}$. 
In this scenario, the spontaneous breaking of chiral symmetry can take place at a finite scale $k \gtrsim M_{\rm Planck}$, if $e_{\mathrm{max}}>e_{\mathrm{crit}}$. 
In this case, the fixed point in the four-fermion interactions $\lambda_{\pm}$ vanishes into the complex plane, potentially triggering critical behavior. This scenario is analogous to the situation in QCD, cf.~\cite{Alkofer:2000wg,Gies:2002hq,Braun:2005uj,Gies:2005as,Braun:2006jd,Braun:2006zz,Braun:2008pi,Haas:2010bw,Braun:2010vd,Braun:2011pp,Braun:2011fw}. 
The lower bound on the number of fermions in the asymptotically free case for the gauge coupling is strictly smaller than the bound arising from the interacting fixed point, i.e., $N_{\mathrm{F}}^{\mathrm{crit}}(e_{*}=0)<N_{\mathrm{F}}^{\mathrm{crit}}(e_{*}>0)$.

We highlight a
potential
caveat concerning the direct association of a divergence in the four-fermion couplings with the spontaneous breaking of chiral symmetry.  In principle, other channels might be critical and a condensate linked to the breaking of a different symmetry might be formed. An example in three dimensions can be found, e.g., in 
\cite{Braun:2014wja}. 
 Within,
QED$_3$ 
a
phase 
 dominated by condensate-formation in
the vector-channel 
characterized by Lorentz symmetry-breaking  was found. 

 Within the context of our work, the search for an asymptotically safe fixed point in the chirally symmetric theory space is not affected by similar subtleties regarding condensate-formation. In contrast, the scenario outlined just above, relying on the asymptotically free fixed point, in which \emph{spontaneous} symmetry breaking occurs, would require closer investigation to ensure that condensate formation occurs in the chiral-symmetry-breaking channel.

\subsubsection{
 Critical value of the gauge coupling at finite fermion number}
\label{sec: ecrit}
\begin{figure}[t]
	\centering
	\includegraphics[width=.9\linewidth]{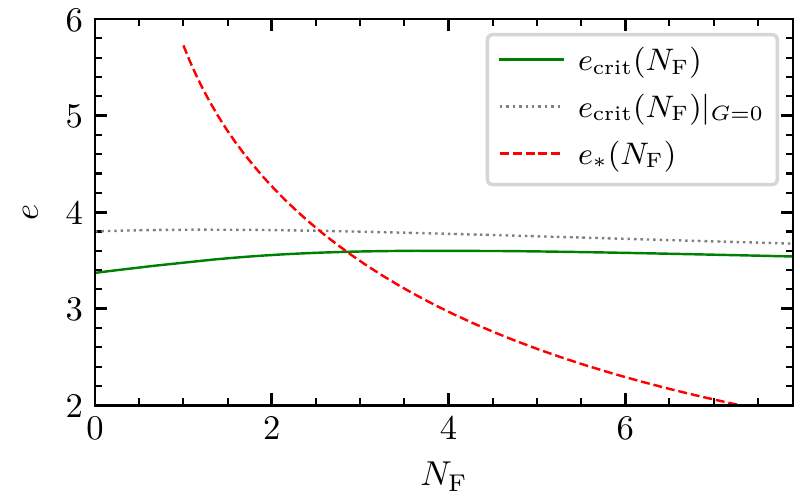}
	\caption{The green solid line shows the critical value of the Abelian gauge coupling $e$ at which the system is driven to criticality, including the impact of asymptotically safe quantum gravity, as a function of the number of Dirac fermions $N_{\mathrm{F}}$. The gray dotted line shows the critical value of $e$ in the absence of quantum gravity, i.e., for $G=\Lambda=0$. The red dashed line shows the fixed-point value $e_{*}$ of the interacting fixed point, which is induced by quantum gravity, cf.~Eq.~\eqref{eq: deshem}.}
	\label{fig: ecrit}
\end{figure}

To understand the dynamical mechanism behind the result in the previous subsection, we now explore the 
 critical value of the gauge coupling, at which the four-fermion fixed points are driven to a collision.
To investigate the corresponding critical value $e_{\mathrm{crit}}$ under the impact of asymptotically safe quantum gravity, we still evaluate $G$ and $\Lambda$ on the extension of the Reuter fixed point as a function of $N_{\mathrm{F}}$, but leave $e$ as a free parameter in a first step. The result is exhibited in Fig.~\ref{fig: ecrit}. There, we compare the critical value of the gauge coupling under the impact of gravity, given by the continuous green line, with the corresponding value without gravity, given by the gray dotted line. Comparing both, it becomes evident that the inclusion of asymptotically safe quantum gravity decreases the critical value, potentially expediting chiral symmetry breaking. This effect results from the interplay of gauge and gravity fluctuations, encoded
in the terms in the terms $\sim e^2G$ in Eq.~\eqref{eq: betaplusminus}, since these terms contribute to the inducing term $c$ in Eq.~\eqref{eq: betaschem}.
 
The red line in Fig.~\ref{fig: ecrit} indicates the value of the interacting, asymptotically safe fixed point $e_*(N_{\mathrm{F}})$ of the Abelian gauge coupling. For $N_{\mathrm{F}}\lesssim2$, the would be fixed-point value is larger than the critical value, indicating that, within our truncation a fixed point with finite value of the Abelian gauge coupling and light fermions cannot be realized. Above $N_{\mathrm{F}}\approx3$, the interacting fixed point $e_*$ lies below the critical value, such that there is no chiral symmetry breaking for three or more Dirac fermions in our truncation. As $N_{\mathrm{F}}$ is increased further, the fixed-point value $e_{*}$ falls even further below the critical line. Thus, if sufficiently many fermions exist, their light nature can be reconciled with a calculable value of the gauge coupling in the IR.

In Fig.~\ref{fig: CSBGLambdaPlot}, we show the boundary between the regions where the fermionic interactions feature real fixed points, and where the chirally symmetric theory-space is UV-incomplete, in the parameterspace spanned by $G$ and $\Lambda$. The pink line indicates the asymptotically safe fixed point as a function of $N_{\mathrm{F}}$, which results in the lower bound on the number of fermions that are necessary for a UV-complete, chirally symmetric, theory.

Our quantitative results are subject to systematic uncertainties related to the choice of truncation. Nevertheless, the dynamical interplay observed here is expected to remain robust, since $e_{*}$ decreases as a function of $N_{\mathrm{F}}$. Thereby, systems with few fermions are more likely to be prone to the onset of criticality than systems with larger numbers of fermions. Whereas the critical fermion number $N_{\mathrm{F}}\approx 3$ should accordingly be understood to come with a systematic uncertainty, the qualitative result, consisting in a \emph{lower bound} on the fermion number, is expected to be robust. 

\subsection{Effective-field-theory setting for quantum gravity}
Let us now take a broader effective-field-theory (EFT) perspective. We still evaluate the Abelian gauge coupling $e$ on the interacting fixed point $e_*$, but treat the gravitational couplings $G$ and $\Lambda$ as input parameters. 
In this way, we consider an EFT treatment of gravity, and investigate the system for a set of initial conditions for $G$ and $\Lambda$, which would be given by a more microscopic theory  holding beyond a scale $k_{\mathrm UV}>M_{\mathrm{Pl}}$. This treatment assumes that a quantum field theoretic description of quantum gravity and matter beyond the Planck scale is still possible but breaks down at $k_{\mathrm{UV}}$, see also \cite{Percacci:2010af,deAlwis:2019aud,Held:2020kze}. Furthermore, we demand that the initial condition for $e(k)$ lies within the basin of attraction of the interacting fixed point $e_*$.
Under those conditions, $e(k)$ will be attracted towards the interacting fixed point $e_*$  under the transplanckian RG flow. The fixed-point-value $e_{*}$, i.e., the solution of Eq.~\ref{eq:betae}, will thus be realized approximately at the Planck scale.

\begin{figure}[t!]
	\centering
	\includegraphics[width=.9\linewidth]{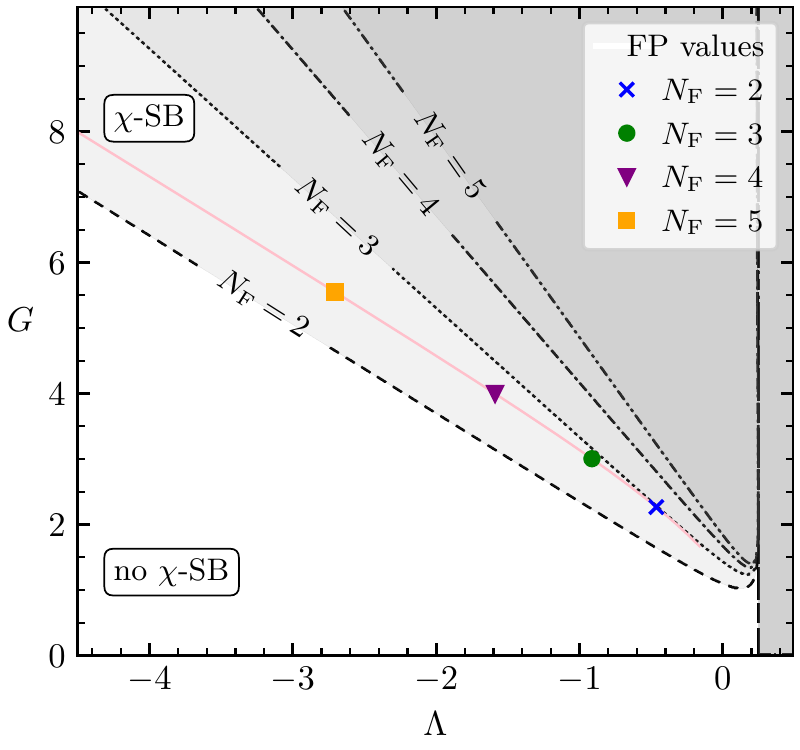}
	\vspace*{-.1cm}
	\caption{For a fixed number of fermions, the dashed (dotted, dashdotted, dashdotdotted) line represents the boundary between the region of intact chiral symmetry (white area) and spontaneously broken chiral symmetry (shaded area). The pink solid line, together with the cross, circle, triangle and square, indicate the gravitational fixed point as a function of the number of fermions.}
	\vspace*{-.2cm}
	\label{fig: CSBGLambdaPlot}
\end{figure}
In Fig.~\ref{fig: CSBGLambdaPlot} we show the boundary between the chirally symmetric (white region) phase and the phase of spontaneous chiral symmetry breaking (gray region) for different values of $N_{\mathrm{F}}$. For any initial values of $G$ and $\Lambda$ situated to the left of the respective line, chiral symmetry will remain unbroken, allowing for light fermions in the IR. If, however, a fundamental theory of quantum gravity gives rise to values of $G$ and $\Lambda$ situated in the respective shaded region, chiral symmetry will be broken spontaneously. As this would be a quantum gravitational effect, the associated mass scale would be associated with the Planck scale, resulting in a tension with the presence of light fermions. 
The area where chiral symmetry is broken decreases as a function of $N_{\mathrm{F}}$. Therefore, more parameter room is available for theories which feature light fermions, if the number of fermions is sufficiently big.

\section{Discussion}

We have investigated the lightness of charged fermions under the impact of quantum gravity. Our results indicate that the existence of a critical value $e_{\mathrm{crit}}$ beyond which chiral symmetry is  broken, persists under the impact of asymptotically safe quantum gravity, cf.~Fig.~\ref{fig: ecrit}. More specifically, we find that the inclusion of quantum gravity fluctuations reduces the critical value $e_{\mathrm{crit}}$, such that chiral symmetry is broken for lower values of $e$ than in the absence of dynamical gravity.

We combine this observation with the $N_{\mathrm{F}}$-dependence of the nonzero fixed-point value for the gauge coupling, which decreases as a function of $N_{\mathrm{F}}$. Light fermions can thus only be accommodated at the fully interacting gauge-gravity fixed point beyond a lower bound on the number of Dirac fermions, $N_{\mathrm{F}}$, which is $N_{\mathrm{F}}\approx 3$ within our approximations.

We highlight that the mechanism of gravitational catalysis in asymptotically safe quantum gravity, which requires a non-vanishing background curvature and is therefore absent in the present setup, already gives rise to a maximal number of fermions that could be allowed for chiral symmetry to remain intact \cite{Gies:2018jnv}. Together with a potential lower bound on the number of fermions, which exists if the interacting fixed point for the Abelian gauge coupling is realized, asymptotically safe quantum gravity might thus only allow for the existence of light fermions within a non-trivial range for the number of fermions.

It is worth pointing out a technical point here:
Our results rely on the background field approximation. However, qualitatively similar results hold with the scale dependence of the Newton coupling and the cosmological constant extracted in a momentum-dependent way in the fluctuation approach, see \cite{Pawlowski:2020qer} for a review, cf.~\cite{Eichhorn:2018ydy}, together with the extraction of the anomalous dimension for the gauge field at finite momentum $p^2=k^2$ \cite{Christiansen:2017cxa}. 
 In brief, the shared qualitative features are i) the existence of a non-trivial lower bound $N_{\mathrm{F, crit}}>1$, ii) unbroken chiral symmetry in the weak-gravity regime and iii) the shift of the boundary of chiral-symmetry breaking towards a more strongly coupled regime, when increasing the number of fermions.

As argued above, the mechanism at the heart of the our lower bound is expected to be qualitatively robust. Yet, a quantitative characterization of the number of light fermions, lies beyond the scope of this work. We stress that a more comprehensive investigation of the interacting fixed point of the Abelian gauge coupling $e_{*}$, requires the inclusion of, e.g.,  $(F_{\mu\nu}F^{\mu\nu})^2$ operators \cite{Christiansen:2017gtg,Eichhorn:2019yzm}. Those operators give rise to a ``weak gravity bound'', see also \cite{Eichhorn:2011pc,Eichhorn:2012va,Eichhorn:2016esv}, since they lead to new divergences in the matter sector, once gravitational fluctuations become too strong. An investigation of the interplay of the "weak-gravity bound" in the Abelian gauge sector, together with the boundary between intact and broken chiral symmetry, might in the future allow for a more quantitative understanding of the impact of asymptotically safe quantum gravity on Standard-Model like matter fields. 

As an additional technical point, a further direction allowing to test the quantitative predictions of our mechanism, is to extract the scale dependence of the gauge coupling $e$ from the three-point vertex of two fermions and one gauge field, instead of using the Ward identity, as in Eq.~\ref{eq: WID}. While the scale dependence extracted in both ways agree at one loop, this is not necessarily the case when investigating an interacting, gravity-generated fixed point, see also \cite{Eichhorn:2017lry}. However, the concept of \textit{effective universality}, as discovered for different avatars of the Newton coupling \cite{Denz:2016qks,Eichhorn:2018akn,Eichhorn:2018ydy,Eichhorn:2018nda}, could also hold for the gauge coupling, which would provide further evidence for the near-perturbative nature of asymptotically safe gravity with matter.\\

\emph{Acknowledgements:}
We thank M. Reichert for discussions.
This work is supported by a research grant (29405) from VILLUM FONDEN.  Additionally, M.~S.~acknowledges support by the German Academic Scholarship Foundation and gratefully acknowledges hospitality at CP3-Origins, University of Southern Denmark, during final stages of this project. GPB is grateful for the support by CAPES under the grant no.~88881.188349/2018-01 and CNPq no.~142049/2016-6 and thanks the ITP at Heidelberg University for hospitality when this project was initiated. We gratefully acknowledge funding from the Heraeus-Stiftung for a research group retreat during which this project was developed.
\bibliography{references,manuals}
\end{document}